\documentclass{article}%
\usepackage{makeidx}
\usepackage{amssymb}
\usepackage{eurosym}
\usepackage{amsfonts}
\usepackage{amsmath}
\usepackage{graphicx}%
\begin{document}

\title{CSAW: A Dynamical Model of Protein Folding}
\author{Kerson Huang\thanks{Email: kerson@mit.edu}\\Physics Department, \\Massachusetts Institute of Technology, \\Cambridge, MA 02139, USA\\and\\Zhou Pei-Yuan Center for Applied Mathematics, \\Tsinghua Univeristy, \\Beijing 100084, China}
\maketitle

\begin{abstract}
CSAW (conditioned self-avoiding walk) is a model of protein folding that
combines the features of SAW (self-avoiding walk) and the Monte-Carlo method.
It simulates the Brownian motion of a chain-molecule in the presence of
interactions. We begin with a simple model that takes into account the
hydrophobic effect and hydrogen bonding. The results show that the hydrophobic
effect alone establishes a tertiary structure, which however has strong
fluctuations. When hydrogen bonding is added, helical structures emerge, and
the tertiary structure becomes more stable. The evolution of the chain
exhibits a rapid hydrophobic collapse into a "molten globule", whose slow
transition to the final equilibrium state mimics experimental data. The model
is designed so that one can add desired features step by step.

\end{abstract}

\section{Introduction}

We propose a model of protein folding, CSAW (conditioned self-avoiding walk),
which is a combination of SAW (self-avoiding walk) and Monte-Carlo. The
general idea is that a protein chain in an aqueous solution undergoes Brownian
motion due to random impacts from water molecules. At the same time, it is
subjects to constraints and organized forces. One the constraint is that
residues on the chain cannot occupy the same position. This is taken into
account through SAW. Other forces and interactions, such as the hydrophobic
effect arising from the water network, and interactions among the residues on
the protein chain, are taken into account through a Monte-Carlo procedure.

Just as simple random walk simulates the Langevin equation, the present
procedure simulates a generalized Langevin equation that describes the protein
chain in a native environment.

Our working goal is to begin with a very simple model that nevertheless
captures the essence of the protein chain, and ad on more realistic features
step by step. The starting point is to construct the backbone of the protein
molecule as a sequence of connected "cranks" made up of the bonds lying in the
amide plane. The orientation of a crank relative to its predecessor is
specified by the usual pair of torsional angles. We can then attach extra
bonds and side chain as desired, like adding components in an erector set.

As a first attempt, we take the hydrophobic effect into account by assigning
each residue\ with a hydrophobic index 0 or 1, and define a configuration
energy that favors the shielding of hydrophobic residues from the environment.
This simple models show that the hydrophobic effect alone establishes a
tertiary structure, which however has strong fluctuations.

Next, we take into account hydrogen bonding among the residues. To do this, we
first\ attach $O$ and $H$ atoms to the cranks, through bonds with the correct
lengths and angles. We then introduce an attractive potential between the $O$
and $H$ atoms, which depends on their separation and the relative angle of
their bonds. The addition leads to the following new phenomena:

\begin{itemize}
\item Secondary structure emerges in the form of alpha-helices, and stabilizes
the tertiary structure.

\item The chain rapidly collapses to an intermediate "molten globule" state,
which lasts for a relatively long time before decaying slowly to the native
state. The evolution of the radius of gyration agrees qualitatively with
experimental observations.
\end{itemize}

These preliminary results indicate that, compared with molecular dynamics, the
CSAW model uses less computer time, and makes the physics more transparent. It
merits further examination and development.

This work was first described in lectures given at Tsinghua University in
2005, which were sequels to an earlier lecture series [1]. In that earlier
series we developed the view that protein folding should be treated as a
stochastic process [2], and CSAW is an implementation of that philosophy.

\section{Random walk and Brownian motion}

Let us review the Brownian motion of a single particle suspended in a medium.
Its position $x(t)$ is a stochastic variable describe by the Langevin equation
[3]%
\begin{equation}
m\ddot{x}=F(t)-\gamma\dot{x}.
\end{equation}
The force on the particle by the medium is split into two parts: the damping
force $-\gamma\dot{x}$ and a random component $F\left(  t\right)  $, which
belongs to a statistical ensemble with the properties%
\begin{align}
\left\langle F(t)\right\rangle  &  =0,\nonumber\\
\left\langle F(t)F(t^{\prime})\right\rangle  &  =c_{0}\delta\left(
t-t^{\prime}\right)  .
\end{align}
where the brackets $\langle\rangle$ denote ensemble average. The equation can
be readily solved. What we want to emphasize is that It can also be simulated
by random walk [4]. Both procedures give rise to diffusion, with diffusion
coefficient $D=\frac{c_{0}}{2\gamma^{2}}$.

In the presence of a regular (non-random) external force $G(x)$, the Langevin
equation may not be soluble analytically , but we can solve it on a computer
via \textit{conditioned random walk. }We first generate a trial step at
random, and accept it with a probability according to the Monte-Carlo method,
which is usually implemented through the Metropolis algorithm [5]. Let $E$ be
the energy associated with the external force $G$, In this case, $E$ is just
the potential energy such that $G=-\partial E/\partial x$. Let $\Delta E$ be
the energy change in the proposed update. The algorithm is as follows:

\begin{itemize}
\item If $\Delta E\leq0,$ accept the proposed update;

\item If $\Delta E>0,$ accept the proposed update with probability
$\exp\left(  -\Delta E/k_{B}T\right)  .$
\end{itemize}

\noindent The last condition simulates thermal fluctuations. After a
sufficiently large number of updates, the sequence of state generated will
yield a Maxwell-Boltzmann distribution with potential energy $E$.

The conditioned random walk may be described in terms of the Langevin equation
as follows:%
\begin{equation}
m\ddot{x}=\underset{\text{Treat via random walk}}{\left[  F(t)-\gamma\dot
{x}\right]  }+\underset{\text{Treat via Monte-Carlo}}{G(x).}%
\end{equation}

\section{SAW and CSAW: Brownian motion of a chain molecule}

A protein molecule in its denatured (unfolded) state can be simulated by a
random coil. That is, it is a long chain undergoing Brownian motion in a
medium. The time development is a sequence of random chain configurations,
with the restriction that two different residues cannot overlap each other. An
instantaneous configuration is therefore a \textit{self-avoiding walk} (SAW)
--- a random walk that never crosses itself.

For illustration, we can generate a SAW with hard-sphere exclusion on a
computer, as follows:

\begin{itemize}
\item Specify a step size, a starting position, and a radius of exclusion
around an occupied position.

\item * Make a random step of the given magnitude and arbitrary direction.

\item If the step does not overlap any of the previous positions, accept it,
otherwise go to *.
\end{itemize}

\noindent By repeating the procedure $N$ times, we generate a SAW representing
the instantaneous configuration of an $N$-chain.

A uniform ensemble of chains can be generated by the pivoting method [6] [7]:

\begin{itemize}
\item Start with an initial SAW chain. Take one end as a fixed origin.

\item ** Choose a random position on the chain. Rotate the end portion of the
chain about this position, through a randomly chosen rotation.

\item If the resulting chain does not overlap the original chain, accept it as
an update, otherwise go to **.
\end{itemize}

\noindent The rotation is a random operation making use of the allowed degrees
freedom of the chain. In a protein chain, this consists of rotation the two
torsion angles.

After a sufficient number of warm-ups, the procedure generates a sequence of
SAW chains belonging to a statistical ensemble that is ergodic and uniform,
\textit{i.e}., it includes all possible SAW's with equal probability. The
sequence of updates may be taken as a simulation of the approach to thermal
equilibrium with the environment.

Formally speaking, we are simulating the solution of a generalized Langevin
equation of the form [8]%

\begin{equation}
m_{k}\mathbf{\ddot{x}}_{k}=\mathbf{F}_{k}(t)-\gamma_{k}\mathbf{\dot{x}+U}%
_{k}(\mathbf{x}_{1}\cdots\mathbf{x}_{N})\text{,\ \ \ \ (}k=1\cdots
N)\text{\ \ \ \ \ \ \ \ \ \ \ } \label{Langevin-chain}%
\end{equation}
where the subscript $k=1\cdots N$ \ labels the residue along the chain. The
term $\mathbf{U}_{k}$ includes the forces that maintain a fixed distance
between successive residues, and prohibit the residues from overlapping one
another. Other interactions are neglected when the chain is in an unfolded
state. The SAW algorithm is a special case of a Monte-Carlo procedure in which
the energy change is either 0 (if there are no overlaps), or $\infty$ (if
there is an overlap).

The protein begins to fold when placed in an aqueous solution, because of the
hydrophobic effect. To model the process, we must include the interactions
among residues, as well as their interactions with the medium. The generalized
Langevin equation still has the form (\ref{Langevin-chain}), with added
interactions:%
\begin{equation}
m_{k}\mathbf{\ddot{x}}_{k}=\,\underset{\text{ \ \ Treat via SAW}}{\left(
\mathbf{F}_{k}-\gamma_{k}\mathbf{\dot{x}+U}_{k}\right)  }+\underset
{\text{Treat via Monte-Carlo}}{\mathbf{G}_{k}.}%
\end{equation}
As indicated on the right side of the equation, we simulate this equation via
CSAW (\textit{conditioned self-avoiding walk}). The interaction forces
$\mathbf{G}_{k},$ which was neglected in the unfolded case, is now taken into
account through Monte-Carlo.

To learn how to include the forces $\mathbf{G}_{k}$. we first consider simple
examples, and then try to build a more realistic model step by step.

\section{The action of water}

The hydrophobic effect drives a protein chain to its folded configuration, and
maintains it in that configuration. The effect has its origin in the hydrogen
bonding between water molecules results in a fluctuating water network with a
characteristic time of 10$^{-12}$s\ .

The residues in a protein molecule can be chosen from a pool of 20 amino
acids, differing from each other only in the structure of the side chain,
which can be polar (possessing an electric dipole moment) or nonpolar. The
polar ones can form hydrogen bonds with water, and are said to be
\textit{hydrophilic}. The nonpolar ones cannot form hydrogen bonds, and are
said to be \textit{hydrophobic}. The presence of the latter robs waters
molecules of the opportunity to from hydrogen bonds, and disrupts the water
network. \ It induces an effective pressure from the water network, to have
the hydrophobic residues shielded by hydrophilic ones, to prevent them from
coming into contact with water. The folded state of the protein is determined
by how this could be done with least cost to everyone involved.

While the side chains may be hydrophilic or hydrophobic, the backbone is
hydrophilic. Frustration arises when one attempts to bury the hydrophobic
residues in the interior of the molecule, because the part of backbone that
gets buried loses contact with water. The frustration is resolved by the
formation of secondary structures, $\alpha$-helices and $\beta$-strands, in
which hydrogen bonding occurs among different residues. Thus, the tertiary
structure and the secondary structure are coupled to each other.

The action of water on the protein molecule can be divided into 3 aspects:

\begin{itemize}
\item Thermal motion of water molecules induce Brownian motion of the protein
chain, folded or not. This is the thermalizing interaction.

\item The water network applies pressure to the protein, squeezing it into the
folded state and maintain it in that state. This is the hydrophobic effect.

\item The water network has vibrational modes in a wide range of frequencies.
The low-frequency modes can resonate with low normal modes of the protein
molecule. These are associated with shape oscillations of the protein, and the
input energy is transferred to modes of higher frequency, initiating an energy
cascade down the length scales of the protein molecule. The significance of
this interaction was explored [9], but remains to be understood.
\end{itemize}

The resonance between water and protein can be illustrated in more detail.
Fig.1 shows both the spectrum of water [10] and the spectrum of a small
protein [11], both based on calculations. We can see there is a prominent peak
in both spectrum at a frequency of about 100 cm$^{-1}.$ Resonant oscillation
between protein and water can occur at this peak. In water, this mode
corresponds to a collective motion of the $O$ atoms in the network, while In
the protein it corresponds to a shape oscillation. 

\section{First model}

In the first model, we take the residues as impenetrable hard sphere, with
diameter equal to the distance between successive residues. The hydrophobic
effect is taken into account using a simple "PH model" [12] in which the
residues are either polar (P) or hydrophobic (H).

\begin{itemize}
\item The centers of the hard spheres (corresponding to C$_{\alpha}$ atoms,)
are connected by "cranks" make up of bonds. to be described in detail later.
The relative orientation of two successive cranks is specified by the two
torsional angles $\psi,\phi$. With one end fixed, the degrees of freedom of an
$N$-chain are the $N-1$ sets of angles $\{\psi_{j},\phi_{j}\}$ $\left(
j=2,\cdots,N\right)  .$

\item The energy $E$ of a configuration is given by%
\begin{align}
E  &  =-gK,\nonumber\\
K  &  =\text{Contact number of hydrophobic residues.} \label{energy}%
\end{align}
A residue's contact number is the number of other residues in contact with it,
\textit{not counting those lying next to it along the chain}. Two residues are
in contact when their centers are separated by less than a set fraction (say,
.1.2) of the residue diameter.

\item When two H are in contact, the total contact number increases by 2. This
induces an attractive force between H residues.
\end{itemize}

Remarks:

\begin{itemize}
\item The denatured chain in a non-aqueous solution corresponds to $g=0.$

\item Only $E/(k_{B}T)\ $appears in the Metropolis algorithm, Thus only the
combination $g/(k_{B}T)$ matters.
\end{itemize}

\section{Example in 2D}

We first illustrate the model in a simple setting to show that it has merit.
Here, the monomer on the chain are connected by straight bonds.

Consider a chain of 7 circular monomers in 2D continuous space, with only one
H. The configuration for maximal shielding of H from the medium is a hexagonal
close pack with H at the center. We shall show that CSAW reproduces this result.

Take the hard sphere diameter to be $1$. The H monomer is considered shielded
by another monomer, if the distance to the latter is less than $1.2$. The
maximum number of nearest neighbors is $4,$ as illustrated in the Fig.2.

Fig.3 displays the evolution of the 2D chain in different runs, with H (the
black dot) placed at different positions on the chain. The numerals under each
configuration are the Monte-Carlo steps. We see that all runs tend to the
hexagonal close pack, thus illustrating "convergent evolution". The lowest
panels shows the evolution of the rms radius. 

\section{The protein chain}

We review salient properties of the protein chain, with a view to design a
model that can be incrementally improved as work progresses.

Fig.4 shows a schematic representation of the protein chain. The center of
each residue is a carbon atom denoted $C_{\alpha}$. Of particular relevance is
whether a molecular group in a residue is polar (P, or hydrophilic) or
nonpolar (H, or hydrophobic). The backbone contains polar subunits $N-H$ and
$C=O$ that can participate in hydrogen bonding. The former wants to donate an
$H$ atom, while the latter seeks an $H$ atom:%
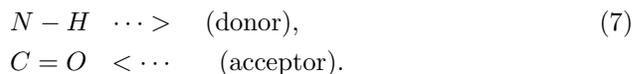
\begin{align}
&  N-H\text{\ \ }\cdots>\text{ \ \ (donor),}\\
&  C=O\text{ \ }<\cdots\text{ \ \ \ \ (acceptor).}\nonumber
\end{align}
They cannot bond with each other, however, because the bond length is not
correct. They may bond with water molecules in the medium, or with other
residues. The latter possibility leads to the secondary structures in the
protein chain. The side chains $R$ attached to a $C_{\alpha}$ atom, may be P
or H, and this determines whether the residue is taken to be P or H.

Geometrically, the C$_{\alpha}$ atom sits at the center of a tetrahedron form
by the four atoms NRHC, as shown in Fig.5. The $H$ atom here does not
participate in hydrogen bonding, because the $C_{\alpha}-H$ group is nonpolar.

The connecting group between two successive $C_{\alpha}$'s is made up of four
atoms $OCNH$, which lie in a single plane, the \textit{amide plane}. The bonds
in $-C-N-$ are arranged in the shape of a "crank", as illustrated in the upper
panel of Fig.6. The bond lengths and angles of the crank are given in
Table(\ref{TableKey}). As indicated in the lower panel of Fig.6, the backbone
of the protein is a sequence of cranks\ joined at a tetrahedral angle
$\cos^{-1}(-1/3)\approx109.5^{\circ}.$\medskip%

\begin{table}[tbp] \centering
\begin{tabular}
[c]{ll|ll}%
Bonds & (A) & Angles & $(\circ)$\\\hline
$C_{\alpha}-C$ & 1.53 & $C_{\alpha}-C-O$ & 121\\
$C-O$ & 1.25 & $C_{\alpha}-C-N$ & 114\\
$C-N$ & 1.32 & $O-C-N$ & 125\\
$N-H$ & 1.00 & $C-N-H$ & 123\\
$N-C_{\alpha}$ & 1.41 & $C-N-C_{\alpha}$ & 123
\end{tabular}
\caption{Bond lengths and angles in the crank}\label{TableKey}%
\end{table}%

\section{Torsional angles}

The relative orientation of two successive cranks is specified by two
torsional angles $\{\psi,\phi\}.$ In Fig.6, the angles associated with crank 2
are defined as follows:
\begin{align}
\psi &  =\text{Rotation angle of }N_{2}\text{ about axis }C_{\alpha
2}\rightarrow C_{2}.\nonumber\\
\phi &  =\text{Rotation angle of }C_{1}\text{ about axis }C_{\alpha
2}\rightarrow N_{1},\nonumber\\
&  =\text{Rotation angle of }C_{2}\text{ about axis }N_{1}\rightarrow
C_{\alpha2}.
\end{align}
where the subscripts 1,2 refer respectively the atoms in crank 1 and crank 2.

In the flat configuration we have $\psi=\phi=\pi$ by definition. This
configuration, however, cannot be realized, because of a collision between $H$
and $O$ in successive residues. Because of such exclusions, the angles are
restricted to allowed regions, as displayed in the Ramachandran plots shown in
Fig.7. 

A secondary structure is formed by a sequence of cranks all having the same
torsional angles. The three allowed areas on the Ramachandran plot are
associated respectively with the right-handed $\alpha$-helix, the $\beta
$-strand, and the left-handed $\alpha$-helix. Examples are shown in Fig.8. 

A right-handed $\alpha$-helix is formed by setting all angles to the values
$\psi\approx-50^{\circ}$, $\phi\approx-60^{\circ}.$The hydrogen bonding has
the pattern%
\begin{align}
&  (C=O)_{1}\text{ }\cdots(H-N)_{4},\nonumber\\
&  (C=O)_{2}\text{ }\cdots\text{ }(H-N)_{5}.
\end{align}
\textit{etc.}, where the subscripts labels the residue. \ In a $\beta$-strand,
the angles have the approximate values $\psi\approx\phi\approx-150^{\circ}$.
Such strands are matted together to form a $\beta$-sheet, through the
formation of hydrogen bonds.

The secondary structure are unstable in the denatured state, because the
energy advantage in hydrogen bonding does not overcome the desire for more
entropy. In this state, hydrogen bonding switches from residue to residue, or
to water. The characteristic time should be 10$^{-12}$s, if we can take the
water network as reference.

\section{Hydrophobic collapse}

In this exercise, we leave out all interactions except the hydrophobic effect.
The residues are taken be hard spheres with diameter equal to the distance to
the next neighbor, connected to each other by cranks. Thus, we have a chain of
beads that can twist and turn through changes in the torsional angles. Some
computer-generated chains are shown in Fig.9.

Fig.10 shows the evolution of a chain of 10 cranks, with only one H, in three
separate runs with different placements of the\ H. After about 1000
Monte-Carlo steps, the chain collapses to compact configurations with the H
surrounded. A hexagonal geometry can be discerned. although there is
considerable fluctuation in the shape. Close-packing of free spheres in 3D
will lead to hexagonal or cubic close pack, in which each sphere touches 12
nearest neighbors [13]. This cannot be realized here, because of the
intervention of the connecting cranks.

Next, we consider 20 cranks with two H, and the results are displayed in
Fig.11. We can see the that the two H are attracted to each other, and the
other residues try to a close pack with the pair in the center.

Finally we consider a chain of 30 cranks, with 1/3 of the residues being H.
This is the approximate ratio in a real protein. As shown in Fig.12, the chain
settles into a tertiary structure resembling a three-leaf clover. The lowest
panel of the figure shows backbone images of the tertiary structure by RasMol
[14]. The local structure fluctuates considerably, but the general topology is
unmistakable. We conclude that hydrophobic forces alone leads to a
well-defined tertiary structure in the final state.

\section{Hydrogen bonding}

We now improve the model by taking into account hydrogen bonding. First we
must attach the $O$ and $N$ atoms to the crank, so that there are now $C=O$
and $N-H$ bonds in the amide plane. A hydrogen bond is formed between $O$ and
$H$ $\ $from different residues, when

\begin{itemize}
\item the distance between $O$ and $H$ is 2 A, within given tolerance;

\item The bonds $C=O$ and $N-H$ are antiparallel, within given tolerance.
\end{itemize}

\noindent When these conditions are fulfilled, the energy of the configuration
is lowered by the bond energy [15]. The formation of such a bond is
illustrated in Fig.13. The residues are still either hydrophobic (H) of
hydrophilic (P); but they are no longer treated as hard spheres. The
self-avoidance condition now takes into account collisions among the $H$ and
$O$ atoms on the newly added bonds.

The energy of a configuration that replaces (\ref{energy}) is%
\begin{align}
E  &  =-g_{1}K_{1}-g_{2}K_{2},\nonumber\\
K_{1}  &  =\text{No. of nearest neighbors of hydrophobic residues,}\nonumber\\
K_{2}  &  =\text{ No. of hydrogen bonds.}%
\end{align}
In the Monte-Carlo update only the combinations $g_{1}/k_{B}T$ and
$g_{2}/k_{B}T$ are relevant. In practice, we set $k_{B}T=1$ and treat $g_{1}$
and $g_{2}$ as adjustable parameters. At this stage of modeling, we are not
ready to calibrate these parameters through comparison with experiments.

There is an implicit assumption about hydrogen bonding, namely, atoms on the
side chains can bond with water, while atoms on the backbone can only bond
with each other. This is reflected in the division of the energy into the
$g_{1}$ and $g_{2}$ terms. The $g_{1}$ term simulates the hydrophobic effect
effect arising from bonding between side chains and water. The $g_{2}$ term,
on the other hand, counts only the hydrogen bond among the residues
themselves. This division correspond to experimental fact, but by assuming it
we have introduced a bias into the model. We think it does not bias the
results, but only further study can confirm this.

As a test of the hydrogen-bonding interaction, we turn off the hydrophobic
interaction by setting $g_{1}=0.$ With $n=30$, $g_{2}=1,$ an $\alpha$-helix
emerges after 1000 Monte-Carlo steps, as depicted in Fig.14.

\section{Formation of secondary structure}

To illustrate the formation of secondary structure, we consider a chain of 30
residues with 10 hydrophobic residues, at positions 1,3,6,7,10,16,20,23,26,29.
We set $g_{1}=10$, $g_{2}=8$.

The evolution of the radius of gyration is shown in Fig.15. After a rapid
collapse, there is a long stretch that we identify with the molten globule
state, which eventually makes a transition to the native state. The change in
radius is relatively small, and discernible only in a magnified view. The slow
transition to the native state mimics experimental data from the folding of
apomyoglobin [16], as displayed in the lowest panel of the figure.

It is interesting that, during the fast collapse, the radius first decreases,
and then expands before resuming the collapse. This could be an example of the
"Reynolds dilatancy" observed in granular flow [17].

Images of the chain are shown in Fig.16. The ones at 2000 and 40000 steps
correspond to the molten globule, and the last one at 60000 corresponds to (as
far as we can tell) the native state. The all show two short helices connected
by a loop, with no significant differences to the eye. \ That is, the molten
globule already possesses the architecture and the compactness of the native state.

Other runs with the same parameters do not yield the same transition points
between the molten globule and the native state. This suggests that the
transition is initiated by nucleation, as in a first-order phase transition. A
more careful statistical study is in progress.

\section{Conclusion}

The advantage of CSAW is that it is simple in conception, flexible in design,
and fast in execution. We can start with a backbone make up of a sequence of
bare cranks, and add components step by step. At each stage we can see the
difference made, thereby gain insight into the physical roles played by
different elements in the chain. Our eventual goal, which does not appear to
be extravagant, is to model a protein molecule with full side chains.

I thank Zuoqiang Shi, physics graduate student at Tsinghua University, for
helping me with the computations on the secondary structures.

\newpage

{\LARGE References}

[1] K. Huang, \textit{Lectures on Statistical Physics and Protein Folding}
(World Scientific Publishing, Singapore, 2005), hereafter referred to as
\textit{Lectures}.

[2] K. Huang, \textit{Introduction to Statistical Physics} (Taylor \& Francis,
London, 2001), hereafter referred to as \textit{Statistical Physics}.

[3] \textit{Lectures}, Chap.10 and Appendix.

[4] \textit{Lectures}, Chap.9; \textit{Statistical Physics}, p.72.

[5] \textit{Statistical Physics}, Sec.18.9.

[6] B. Li, N. Madras, and A.D. Sokal, \textit{J. Stat. Phys}. \textbf{80}, 661 (1995).

[7] T. Kennedy, \textit{J. Stat. Phys. }\textbf{106}, 407 (2002).

[8] V.N. Prokrovskii, \textit{The Mesoscopic Theory of Polymer Dynamics}
(Kluwer Academic Publishers, Dordrecht, 2000).

[9] \textit{Lectures}, Chap.16 and Appendix.

[10] S. Saito, I. Ohmine, \textit{J. Chem. Phys}. \textbf{102}, 3566-3579 (1995).

[11] N. Go, T. Noguti, T. Nishikawa, \textit{Proc. Nat. Acad. Sci., USA},
\textbf{80}, 3696-3700 (1983).

[12] H. Li, C. Tang, N.N. Wingren, \textit{Proc. Nat. Acad. Sci., USA},
\textbf{95}, 4987-4990 (1998).

[13] J. H. Conway and N.J.A. Sloane, \textit{Sphere Packings, Lattices, and
Groups}, 2nd ed. (Springer-Verlag, New York, 1993).

[14] RasMol homepage: http://www.umass.edu/microbio/resmol/

[15] A. V. Smith and C.K. Hall, \textit{Proteins}, \textbf{44}, 344-360 (2001)
uses a similar model of hydrogen bonds in a molecular-dynamics computation.

[16] T. Uzawa, S. Akiyama, T. Kimura, S. Takahashi, K. Ishimori, I. Morishima,
T. Fujisawa, \textit{Proc. Nat. Acad. Sci., USA}, \textbf{101}, 1171-1176 (2004).

[17] O. Reynolds, "On the dilatancy of media composed of rigid particles in
contact. With experimental illustrations", \textit{Philosophical Magazine}
(December, 1885).

\newpage

{\LARGE Figure Captions}

Fig.1. Calculated spectrum of normal modes of liquid water network (upper
panel) and that of a protein molecule (lower panel). They resonate at a
frequency of about 100 cm$^{-1}$.

Fig.2. The hydrophobic residue on the chain is represented by the dark circle.
It can have a maximum of 4 nearest neighbors, not counting those on the chain.

Fig.3. Upper panels: Evolution of the chain, with the hyrophobic residue shown
as a black dot. Regardless of its placemen, the chain tends to a hexagonal
close pack, with hydrophobic residue in the center. Lowest panel: Evolution of
the rms radius, showing hydrophobic collapse and steady state equilibrium.

Fig.4. Schematic representation of the protein chain.

Fig.5. Tetrahedron with the $C_{\alpha}$ atom at the center.

Fig.6. Upper panel: Two successive $C_{\alpha}$ atoms are joined by chemical
bonds in the shape of a "crank". Side chains and H atoms attached to
$C_{\alpha}$ have been omitted for clarity. Lowe panel: The protein backbone
is a sequence of cranks joined at a fixed angle. The orientation of the plane
of one crank, relative to the preceding one, are specified by two torsional
angles $\phi$,$\varphi$.

Fig.7. Ramachandran plots of allowed regions for the torsional angles.

Fig.8. Examples of secondary structure. Hydrogen bonds are indicated by dotted lines.

Fig.9. Examples of a chain of hard-spheres connected by cranks.

Fig.10. Chain of 10 cranks with 1 H, which is represented by the solid circle.
Three different runs are shown, with differrent placements of H. All result in
the H being surrounded by other residues in an attempt to achieve hexagonal or
cubic close packing. \ For clarity, we hide the canks, and connect the
$C_{\alpha}$ atoms by straight bonds.

Fig.11. The hydrophobic effect induces an effective attraction between to
hydrophobic residues.

Fig.12. Chain of 30 cranks, with 1/3 of the residues being H. Lowest panel
displays RasMoL backbone images, at the Monte-Carlo steps as labeled.

Fig.13. A hydrogen bond is formed when the distance between $O$ and $H$ lie
within given limits, and the bonds they belong to are antiparallel within
given tolerance.

Fig.14. Testing the hydrogen-bonding interaction with $g_{1}=0$ (no
hydrophbobic effect). An $\alpha$-helix emerges after 1000 Monte-Carlo steps,
as shown here in backbone and spacfill representations by RasMol.

Fig.15. The strong radius fluctuation during the initial colllapse suggests
"Reynolds dilatancy", a phenomenon in granular flow. After the collapse, the
radius remains constant for a relatively long time. This corresponds to the
"molten globule" stage, which slowly decays to the native state, with a small
decrease of the radius. For comparison, the lowest panel shows experimental
data from the folding of apomyoglobin.

Fig.16. RasMol backbone images of the chain. There are two short helices
connected by a loop. The first two images corresponds to the molten globule
state, and the last corresponds to the native state.

\end{document}